\documentclass[10pt]{article}
\usepackage[latin1]{inputenc}
\usepackage{amsmath}
\usepackage{amsfonts}
\usepackage{amssymb}
\usepackage{graphicx}
\usepackage{geometry}
\usepackage[english]{babel}
\usepackage[nottoc]{tocbibind}

\title{Information Content of a Phylogenetic Tree in a Data Matrix}
\author{Tania Roy, Hsieh Fushing\footnote{Correspondence: Hsieh Fushing, University of California at Davis, CA
95616. E-mail: fhsieh@ucdavis.edu}\\
Department of Statistics, University of California, Davis,\\
Xunde Li, Brenda McCowan and Rob Atwill,\\
Department of Population Health and Reproduction,\\
School of Veterinary Medicine, University of California, Davis.}
\date{}
\begin{document}
	\maketitle
	\begin{abstract}
Phylogenetic trees in genetics and biology in general are all binary. We make an attempt to answer one fundamental question: Is such binary branching from the coarsest to the finest scales sustained by data? We convert this question into an equivalent one: where is the structural information of tree in a data matrix? Results from this conceptual as well as computing issue afford us to conclude a negative answer: Each branch being split into two at each inter-node of tree from the top to bottom levels is a man-made structure. The data-driven computing paradigm Data Mechanics is employed here to reveal that information of tree is composed of a set of selected temperatures (or scales), each of which has a clustering composition strictly regulated by a temperature-specific cluster-sharing probability matrix. The resultant Data Cloud Geometry (DCG) tree on the space of species is proposed as the authentic structure contained in data. Particularly each core clusters on the finest scale, the bottom level, of DCG tree should not be further partitioned because of uniformity. Beyond the finest scale, the branching of DCG tree is primarily based on probability, which induces an Ultrametric satisfying super triangular inequality property. This Ultrametric property differentiates DCG tree from all popular trees based on Hierarchical clustering (HC) algorithm, which typically employs an empirical, often ad hoc distance measure. Since this measure is regulated by the triangular inequality, it is not capable of producing a ``flat" branch, in which all its members (more than two) have equal distances to each others. We demonstrate such information content on an illustrative zoo data first, and then on two genomic  data.
	\end{abstract}

\section{Introduction}
In most scientific fields, particularly in biology, the primary tree structure derived from data is binary: upon each inter-node of tree, a subset of leaves (species) is split or bifurcated further into two smaller subsets. Such a binary splitting procedure only stops when only one single leaf is left at an inter-node. Such kind of binary tree is typically resulted from applying various distance based or character based (\cite{Felsenstein},\cite{Farris70},\cite{Farris83},\cite{Larget}) methods.

Most used distance based methods are Neighbor Joining \cite{SaitouNei} and UPGMA(Unweighted Pair Group Method with Arithmetic Mean) trees \cite{sokal58} , which is nothing but a hierarchical clustering(HC) tree made with ``average" linkage. When HC algorithm is implemented, a distance matrix among all leaves has to be built based on an empirical, most often ad hoc, choice of distance measure. An ad hoc choice of metric is likely real-valued and satisfies triangular inequality, such as Euclidean distance. Three underlying reasons are tangled together behind the man-made bifurcations from the coarsest to the finest scales: the first one is the design of HC algorithm; the second one is the real-valued distance measure, and the third one is the triangular inequality.

The design of HC algorithm is to build a binary tree in a bottom-up fashion. At each tree level, two entities, which could be either one leaf or a subset of leaves, having the smallest distance are pooled together into one bigger subset of leaves.  The concept of distance between two sets of leaves is regulated by a choice of so-called module, such as single-linkage via minimum, complete via maximum, UPGMA via averaging and many others (\cite{slink},\cite{complete}). So when a resultant tree is seen from top-to-bottom, this design of pooling two entities into one bottom-up becomes a bifurcating, or branching one top-down.

Secondly, this capability resulting a strictly bifurcating tree, also called complete tree, is due to the fact that all entries of the distance matrix are distinct real-values. This unintended reason is realized as if the measurement of distance is equipped with an infinite precision. The very concept of a finite precision only leading a limited amount of information contained in data, disregarding its size, is lost explicitly when using the HC algorithm to build a tree in any science. Thirdly, the triangular inequality prohibits the fact all members are simultaneously the center of a set of four or more. In contrast, the super triangular inequality under an Ultrametric allows this to happen. As would be seen below, a distance induced from probability of cluster-sharing is naturally Ultrametric. We first make use one simple concept of uniformity, and then illustrate the aforementioned three facts through an 1D data set here.

The simple concept is: any set of one dimensional (1D) data points observed from a uniform distribution on an finite interval, say $[a, b]$, should constitute a core cluster. It is because, under such a stochastic mechanism, we can not predict anything beyond ``being within $[a, b]$''. In other words, each data point observed from uniform random variable $U[a, b]$ should ideally be ``the one" disguised by pure random noise. That is, each and every data point from $U[a, b]$ should be a center that is basically ``equally" distant to all others. This concept of ``everyone being a center to all others'' matches  exactly with the notion of a core cluster of a tree. Pictorially speaking, a core cluster of tree is meant to be one single flat-bottom-branch. Then two sets of 1D data points observed from two separated uniform distributions, say $[a, b]$ and $[c, d]$ with $b \leq c$, should constitute two separate core-clusters, that is, a tree has only one level with two flat-bottom-branches.

After linking uniformity with a tree, we can claim that any 1D data set embraces a tree structure. It is obvious for categorical and discrete data sets, but it is not immediate for a real-valued one. For the latter data type, the bottom-level of the tree structure, as well be seen below, in fact corresponds with a very well piecewise linear approximation of its empirical distribution, and subsequently corresponds to a possibly gapped histogram with heterogeneous sizes of bins and gaps. All bins of such a gapped histogram satisfy uniformity, and collectively constitute a tree-level with flat-bottom branches. This notion of tree on an 1D data set confirms that any branching made upon a flat-bottom-branch, or a core cluster, is not supported by data.

\begin{figure}
	\centering
	\includegraphics[scale=0.7]{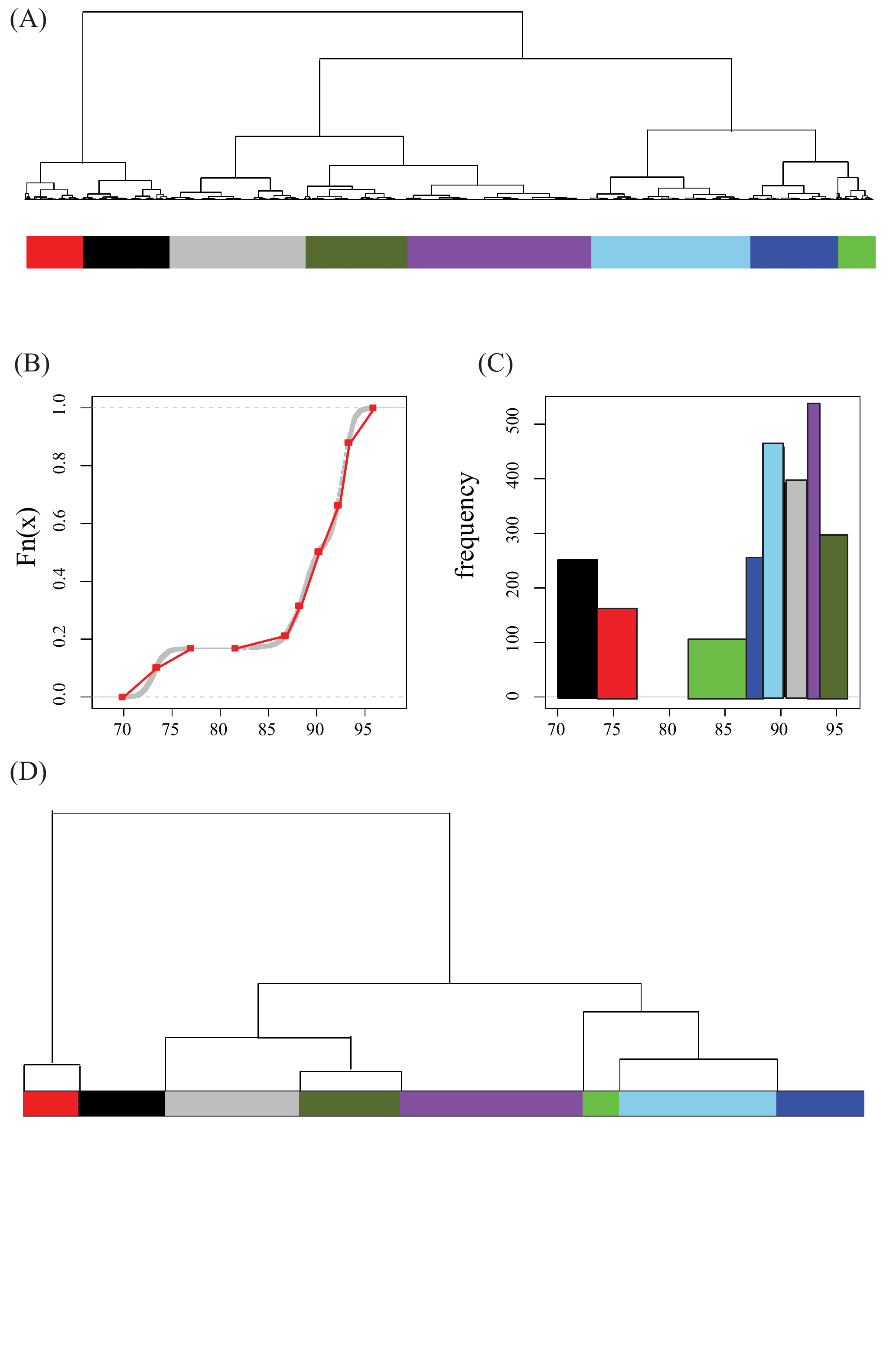}
	\caption{Tree information of 1D data of Kershaw's start-speed: (A) The HC tree; (B) Piecewise linear approximation of the empirical distribution; (C)Color-coded possibly-gaped histogram; (D) Pruned HC tree with branches corresponding to color-coded bins of the histogram in (C).}
	\label{fig:kershaw}
\end{figure}

We illustrate the above arguments through a real 1D data set with more than 3000 real-valued data point. This data set consists of start-speed of all pitches thrown by the Major League Baseball (MLB) pitcher Clayton Kershaw of L.A. Dodgers in 2017 season. When constructing a proper histogram on this data set via Analysis of Histogram (ANOHT) algorithm,  (Hsieh and Roy (2017) \cite{hsiehroy}) a tree structure is revealed in the computing process, as shown through the first three panels of Fig~\ref{fig:kershaw}. The panel (A) is HC tree based on 1D Euclidean distance. For demonstration purpose, a set of 8 branches is chosen to illuminate one reasonable, but not very well piecewise linear approximation in panel (B) and consequently a possibly-gapped histogram in panel (C) with an evident gap. On the 2-branch level, the two black and red colored bins separate start-speeds of cureball (CU) from the start-speeds of all other pitch-types, including fastball (FF), change-up (CH) and slider (SL) in his pitching repertoire. These two bins correspond to two piecewise linear segments that do not approximate their corresponding curve-parts of empirical distribution cure well enough. Therefore the black and red branches in the tree of panel (D) need to be further partitioned according to their tree branching.

In contrast, on the three branch level, the middle branch contains a bifurcated sub-branch, which consists of dark-green and purple colored bins that are occupied by majority of fastball pitches, and an isolated branch, which is gray colored bin that are exclusively occupied by sliders. The right branch also has a bifurcated and an isolated branch. The isolated one colored in light-green is occupied by purely changeup, while the bifurcated one consists of a dark-blue bin for changeups and a light-blue bin for sliders. These 4 out of 6 colored bins, not including dark-green and green bins, individually provide very well piecewise linear approximations their corresponding parts of empirical distribution. They satisfy the uniformity assumption. Thus, any further branching upon any of these 4 bins, as indeed happened in panel (A), is considered not being supported by data.

It is remarked here that ANOHT algorithm is able to construct a gapped histogram with 12 bins, each of which satisfies the uniformity assumption, see Fushing and Roy (2017). That is, the true information of tree in this data should have a bottom-level with 12 flat-bottom branches like that in the panel (D). However, as far as the tree structure beyond the bottom-level is concerned, the tree in panel (D) might not be entirely authentic due to the application of HC algorithm. An alternative to HC algorithm is a clustering algorithm called Data-Could-Geometry (DCG) algorithm, see Fushing and McAssey (2010) and Fushing, et al (2013). The DCG tree is developed and viewed as an minimum energy macrostate in a statistical physics problem. A DCG tree not only has a bottom-level consisting of flat-bottom branches, but also, to our knowledge, has its tree-branching beyond the bottom-level being supported by data. The down side is that DCG requires a bigger computing cost than HC algorithm does.

Next, we consider in general a data set involves with multiple 1D features, that is, each data point (or species) is a multiple dimensional vector. One natural and essential issue facing scientists can be phased as: Where is information of tree supported by the data? Such a data set can be represented by a rectangle matrix format by having all features and species IDs arranged along the column and row axes, respectively. Then the equivalent question becomes: what is the counterpart of a gapped histogram in a 2D matrix data?

We computationally address such an issue in this paper. The guiding principle of our proposed resolution again is to extract and discover uniformity under such a 2D setting. Such uniformity on the matrix-lattice should be seen through mutliscale patterned ``blocks''. Ideally such block-patterns are framed by two mutually coupled (or dependent) clustering trees, which are superimposed upon the matrix's row and column axes, respectively. We propose to compute and discover these two trees via an algorithm called Data Mechanics, see Hsieh and Chen (2014) \cite{hsiehchen}. Data Mechanics is an iterative computing paradigm. It starts with computing one tree on column axis, for instance, then a distance on row vectors is defined by adapting the column tree structure, and then Data Cloud Geometry (DCG) algorithm is applied to construct a tree on the species(row)-axis. Cyclically iterate this procedure between the two axes in order to establish their crucial dependency. While the DCG algorithm specifically carries out data-driven computations to extract two aspects of information of tree: 1) where are reliable tree-levels; 2) what does its clustering composition look like at a tree-level?

It is noted that, before applying Data Mechanics, any original real-valued rectangle matrix has to go through a re-normalization feature-by-feature. One re-normalization is performed with respect a feature's own histogram, see Fushing and Roy (2017). Such a step is to make sure that all features are comparable, so are all subjects. Being comparable is the basis for evaluating ``similarity" among rows and columns. Only when a reasonable measure of similarity or distance is in hand, the clustering results regarding: who is closer to whom, but away from whom, are to be meaningful.

In essence, Data Mechanics basically and simply iteratively permutes rows and columns in order to construct two mutually dependent marginal clustering trees that frame multiscale block patterns such that blocks on the finest scale are being as ``uniform" as possible. This collection of multiscale block patterns is taken as the deterministic structure of information contained in the data matrix, which can be seen as representing a system in statistical physics. While the collection of block-based ``uniformity" is taken as the stochastic structures of information contained in the data. This pair of deterministic and stochastic structures embedded within in a matrix is exactly in the same format as that exhibited through the possibly-gaped histogram in the 1D layout in Hsieh and Roy (2017) \cite{hsiehroy}.

Three pairs of marginal clustering trees are considered and compared: 1) two HC trees via ``average" module (UPGMA); 2) Data-Could-Geometry (DCG) -tree on species axis coupled with a HC tree via complete linkage module; 3) two HC trees via complete-linkage module.  The DCG clustering algorithm extracts the aforementioned two aspects of information of tree by determining two key factors: 1) which series of scales (or temperature) is informative and coherent with respect to data; 2) at each determined temperature, a clustering composition is derived based on a similarity matrix of cluster-sharing probabilities among all species. A DCG clustering tree is derived by synthesizing all selected scale-specific clustering compositions. Therefore such a DCG tree can exhibit the information of tree on species by revealing: first, the number of tree-levels beyond that of DCG is sustained by the data; 2) secondly, any partitioning or bifurcating on its core clusters belonging to its bottom tree level is not sustained by data.

A real zoo data is illustrated throughout our computational developments in the Method section, while two genetic data sets and two linguistic data sets are analyzed in the Results section.

\section{Illustrating zoo-data example:}
Consider the animal kingdom as the real system of interest. The collection of species of animals in
exhibition in a zoo is the species-node space $\mathcal{X}$ . Its cardinality is $100(= n)$, and the 100
animals are subdivided into 7 categories, as given below. Each animal species is observed
with a 15 binary features in regarding to binary categorical characteristics and a categorical feature of
number of legs. This discrete variable of legs is further categorized and coded into 5 binary dummy variables to reflect the 5 status of legs: 0, 2, 4, 6 and 8. So the cardinality of feature-node space is 20. All 20 features are binary. So there is no need for renormalization.

In this example our goal is limited to exploringthe extents, to which the data embraces the
Tree of Life. The animals can be divided into some large known categories, as the following and each class had been labeled by an unique color.
\begin{enumerate}

\item 41 Mammals:(black) aardvark, antelope, bear, boar, buffalo, calf, cavy, cheetah, deer, dolphin, elephant, fruit-bat, giraffe, girl, goat, gorilla, hamster, hare, leopard, lion, lynx, mink, mole, mongoose, opossum, oryx, platypus, polecat, pony, porpoise, puma, pussycat, raccoon, reindeer, seal,
sealion, squirrel, vampire, vole, wallaby,wolf;

\item 20 Birds:(red) chicken, crow, dove, duck, flamingo, gull, hawk, kiwi, lark, ostrich, parakeet, penguin,
pheasant, rhea, skimmer, skua, sparrow, swan, vulture, wren;

\item 8 Reptiles and Amphibians:(green) pitviper, seasnake, slowworm, tortoise, tuatara, frog, newt, toad;
\item 13 Fishes:(blue) bass, carp, catfish, chub, dogfish, haddock, herring, pike, piranha, seahorse, sole, stingray,
tuna;
\item 18 Arthhropods, Mollusa and others:(cyan) flea, gnat, honeybee, housefly, ladybird, moth, termite, wasp, clam, crab, crayfish, lobster, octopus, scorpion, seawasp, slug, starfish, worm.

\end{enumerate}

\subsection{Method}
Our data-driven computing paradigm Data Mechanics iteratively applies its key computing device, Data Cloud Geometry (DCG) algorithm \cite{fushing2010} on the species axis and another clustering algorithm, which can be HC or DCG, on feature axis. The functional utilities of iterative procedure and DCG algorithm are elaborated step-by-step below, while the details are referred to their original publications. The DCG algorithm is a clustering algorithm very distinct from the HC algorithm on many essential ways, even when they are based on the same symmetric distance matrix denoted as ${\cal D}=[d_{ij}]$ with $i, j$ being the $i-$th and $j$-th involving leaf-nodes. However the computing load needed for DCG algorithm is higher than that for HC algorithm.

\paragraph{DCG algorithm:}

\begin{enumerate}

\item[Step-1] With respect to a temperature (or scale) $T$, a similarity matrix is generated as $S^T({\cal D})=[s^T_{ij}]$ with $ s^T_{ij}=e^{\frac{ -d_{ij}}{T}}$.

\item[Step-2] Then each row of $S^T({\cal D})$ is normalized by its row sum. So $S^T({\cal D})$ becomes a transition probability matrix $P^T({\cal D})$.
\item[Step-3] $P^T({\cal D})$ gives rise to a regulated Markov random walk, which starts randomly from a leaf-node, and then removes a leaf-node, when its number of visits by this Markov random walk has gone beyond a threshold. When a leaf-node is removed, the transition matrix $P^T({\cal D})$ is regenerated by deleting the corresponding row and column. This step is designed to keep the Markov random walk from being trapped in a local region of the data cloud.
\item[Step-4] A trajectory of a regulated random walk will give rise to a leaf-node-removal recurrence time series, which is equipped with several spikes, which indicating the random walk has just enters a new, unexplored local region. Therefore, all leaf-nodes removed between two successive spikes are taken as being in the same cluster with respect to the temperature $T$.
\item[Step-5] So a trajectory will give rise to a binary matrix: its $(i, j)$ entry is coded 1 if  the $i-$th and $j$-th leaf-nodes are in the same cluster.  An ensemble of such trajectories will give rise to an ensemble of such cluster-sharing matrices, which is then summarized into a matrix of cluster sharing probability, denoted as $En[P^T({\cal D})]$. (It is noted that this $\tilde{1} \tilde{1}^T- En[P^T({\cal D})]$ is nearly an Ultrametric, which satisfies the super-triangular inequality: $d(x,y) \leq \max{d(x, z), d(y, z)}$.)
\item[Step-6] The number of significantly non-zero eigenvalues of $En[P^T({\cal D})]$ is taken as the number of clusters, say $N(T)$, being present in the temperature scale $T$, while the explicit clustering composition can be extracted by applying HC algorithm, or other clustering algorithm based on $\tilde{1} \tilde{1}^T- En[P^T({\cal D})]$ as a distance matrix. Denote the resultant clustering composition as ${\cal C}_M(T)$, which contains the memberships of the $N(T)$ clusters.

\item[Step-7] Plot $N(T)$ against $T$ (on horizontal axis). We choose a temperature $T_i$ from each leveling-off or constant segment of this plot. (Typically we choose the middle point.) Denote this set of selected temperatures as $\{T_1, ..., T_K\}$ and their corresponding clustering compositions $\{{\cal C}_M(T_1),..., {\cal C}_M(T_K)\}$. This set of clustering compositions are synthesized into a Ultrametric clustering tree. This tree is called Data Could Geometry (DCG) tree, say ${\cal T}$.
	
\end{enumerate}

\begin{figure}[hbtp]
	\centering
	\includegraphics[scale=0.8]{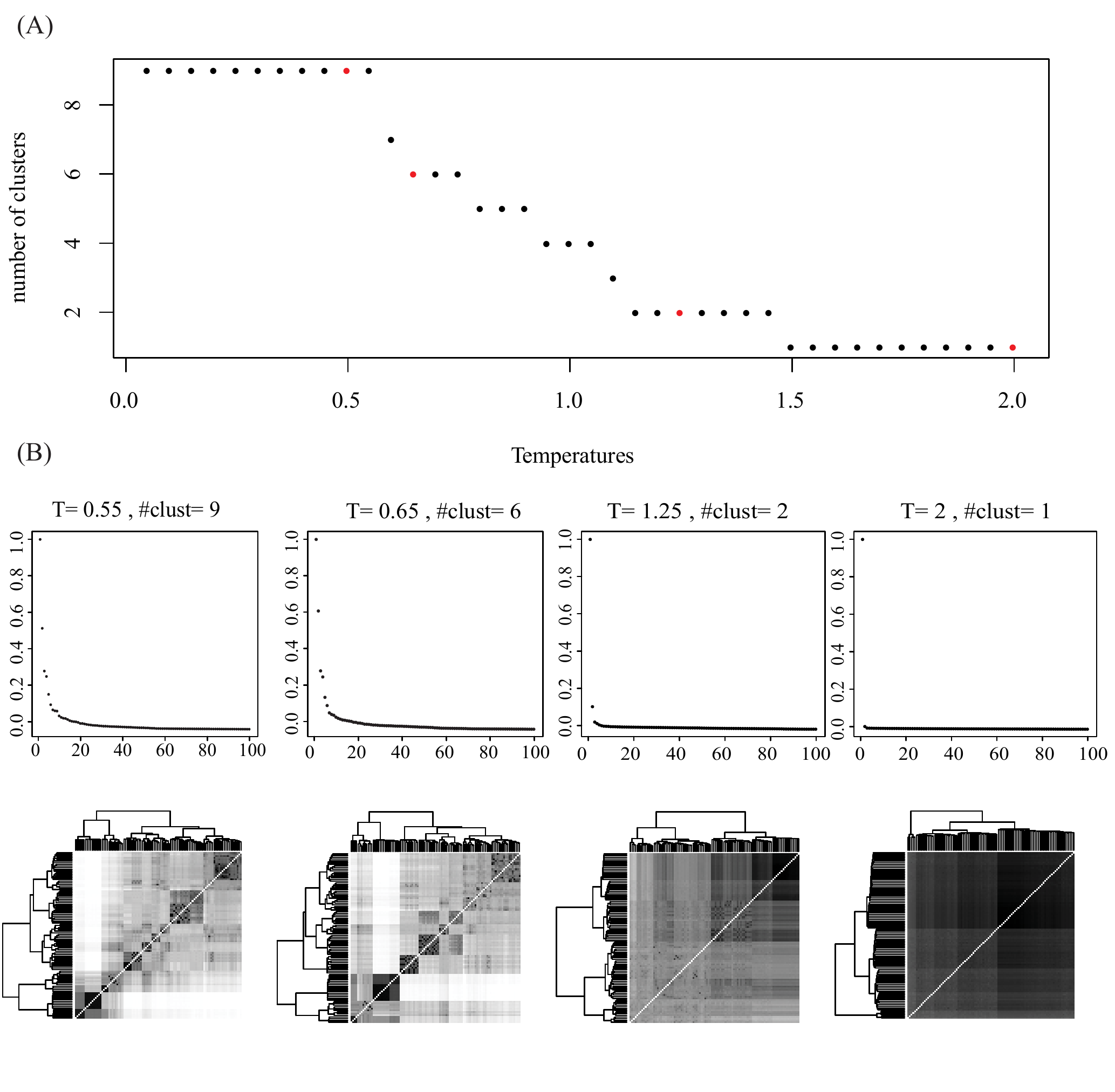}
	\caption{Illustrations of DCG algorithm. (A) Plot of number of clusters vs temperature (Step 7);(B) A series of temperature-specific clustering sharing probability matrices and their eigenvalue-plots.}
	\label{fig:eig-heatmap-zoo}
\end{figure}

The application of DCG algorithm on zoo data are summarized in the two panels of Fig.~\ref{fig:eig-heatmap-zoo}. The panel (A) indicates that there are 6 reliable temperatures or scales at most contained in the data, while the panel (B) brings out that numbers of clusters at four chosen temperatures (marked by red dots) and corresponding clustering compositions on the matrices of probability of cluster-sharing. These data-driven computations in terms of probability are strikingly different from the computations of HC algorithm, which primarily rely on the raw measurements of a ad hoc choice of distance. That is the fundamental reason why such a Ultrametric DCG tree contains authentic information of tree contained in data.


In contrast to DCG algorithm, Data Mechanics algorithm works on a $m \times n$ rectangle matrix, say ${\cal M}$, which could be a re-normalized version from the original data matrix ${\cal M}_0$ with $m$ features measured across all $n$ leaf-subjects.

\paragraph{Data Mechanics:}
\begin{enumerate}
\item[Step-1] We adopt the Euclidean distance measure on all $m$ rows of ${\cal M}$, and construct a distance matrix ${\cal D}^{(R0)}=[d^{(R0)}_{ij}]$. We then apply HC (orDCG) algorithm to build an Ultrametric tree, say ${\cal T}^{(R0)}$ on the row axis.
\item[Step-2] Select a tree level on ${\cal T}^{(R0)}$. Extract the corresponding clustering composition ${\cal C}_M(T^*)$ with $N^*$ clusters. Consider each column being extended with $N^*$ extra dimensions of average among member-rows of clusters of ${\cal C}_M(T^*)$. Define a distance matrix ${\cal D}^{(C1)}=[d^{(C1)}_{I'j'}]$ among the $n$ column with $ d^{(C1)}_{I'j'}$ being calculated as the $m+ N^*$  dimensional Euclidean distance.
\item[Step-3] Based on distance matrix ${\cal D}^{(C1)}$, a DCG tree ${\cal T}^{(C1)}$ is calculated on column axis.
\item[Step-4] Based on one selected level of the tree ${\cal T}^{(C1)}$, an adopted distance measure, say $ d^{(R1)}_{ij}$, on row vectors are devised as in Step 2, and a new distance matrix ${\cal D}^{(R1)}=[d^{(R1)}_{ij}]$ is also computed. We then apply HC (or DCG) algorithm to build a new Ultrametric tree, say ${\cal T}^{(R1)}$ on the row axis.
\item[Step-5] Repeat the Step-2 to Step-4 for two or three times, or until both trees ${\cal T}^{(Rk)}$  and ${\cal T}^{(Ck)}$ are stable.
Let the final two marginal trees are denoted as ${\cal T}^{(R*)}$  and ${\cal T}^{(C*)}$. The final result of Data Mechanics computations is a heatmap described below.
Superimposing the two marginal Ultrametric trees ${\cal T}^{(R*)}$  and ${\cal T}^{(C*)}$ of the row and column axes of ${\cal M}$. These two tree jointly frame the multiscale block patterns on the lattice of ${\cal M}$, which is termed a heatmap of coupling geometry. Ideally all blocks on the finest scale embed with uniformity.

\end{enumerate}


\paragraph{Matrix mimicking:}
In 1-dim setting, simulating data bin-by-bin based on a possibly-gapped histogram computed from a dataset, as shown in panel (D) of Fig~\ref{fig:kershaw}, maybe is the only legitimate way of mimicking original 1D dataset. This mimicking is intended to retain the deterministic and stochastic structures embedded within the data. Mimicking a matrix data has the same goal of retaining structural patterns computed via DM from the original matrix. The multiscale block patterns framed by two DCG trees constitute the deterministic structures, while the composition of uniformity within blocks of the finest scale constitutes the stochastic structure. Thus, the most essential merit of matrix mimicking is to address the issue of whether the finest scale of DCG tree is indeed supported by the data or not.

To achieve this mimicking goal, we need computing algorithms to be able to generate a block, or sub-matrix, subject to a proper collection of constraints inherited from the observed one. This collection of constraints are the observed row-series and column-series of empirical distributions. For a binary block, each empirical distribution constraint is equivalent to a sum. Such constraints in terms of empirical distribution is not only necessary, but also critical for the mimicked matrices to be biological meaningful. This is especially clear when the data matrix is categorical.  Recently matrix mimicking appeared computer science and statistics literatures, such as Bayati (\cite{Bayati}) and Miller (\cite{Miller}), which are designed to fulfill the constraints of row-series and column-series of sums. However our experiences reveal that both mimicking will cost a lot in computing when numbers of rows or columns of a target block (submatrix) go beyond 100. This phenomenon is seen at the genetic data examples reported in Result section.

Since Data Mechanics is able to build the multiscale patterned blocks on a matrix lattice, which becomes a natural platform for divide-and-conquer to work. Hence we adopt Biyaiti's algorithm into an algorithm developed in Fushing, et al. called Binary-Slicing algorithm, see Fushing et al. (2015), to achieve the matrix mimicking goal. Binary-slicing algorithm works by slicing a discrete or digital-coded matrix into a sum of binary matrices.

For a categorical data matrix, the first step was to digitally encode the categories with gap in order to better separate distinct sets of categories. The other approach is to apply the binary-slicing algorithm on each distinct set of categories separately. For instance, it is known that the base pairs $\{A, G \}$ are similar while the base pairs  $\{C, T\}$ are slightly different from the former grup in nature.  Hence according to a common practice $\{A, G, C, T\}$ are coded by $\{1, 2, 5, 6\}$. Next the coded data matrix is permuted according to the data mechanics trees, and seperated into some homogeneous blocks according to the clustering from the two trees. Next a new mimicking algorithm that applies the Binary slicing method in a nested manner was applied on each sub-matrix, to check if they are homogeneous enough. The algorithm for mimicking an  $\{A, G, C, T\}$ matrix is described in the Appendix A in detail. There are two versions of the algorithm described, and the second one being faster was applied for mimicking bigger sub-matrices.

\begin{figure}[hbtp]
	\centering
    \includegraphics[scale=0.5]{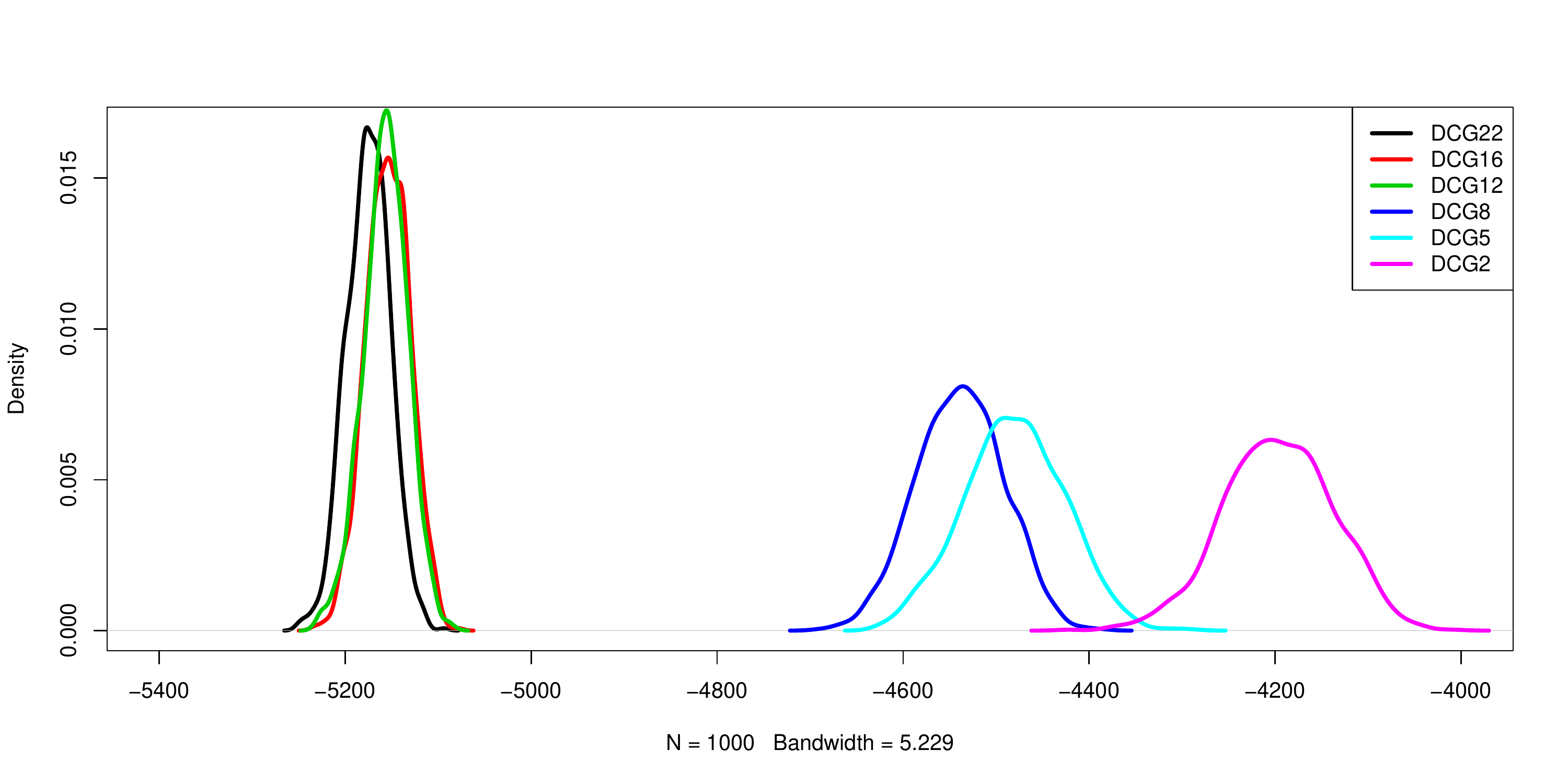}
	\caption{Total variation energy densities from ensembles of mimicked matrices based on DCG tree structure.}
	\label{fig:energyzoocomparison3}
\end{figure}

In the Fig.~\ref{fig:energyzoocomparison3}, we confirm that the DCG tree structure down to its bottom level is supported by the data.

\section{Results:}
\subsection{Zoo data} 
\begin{figure}[hbtp]
	\centering
	\includegraphics[scale=0.7]{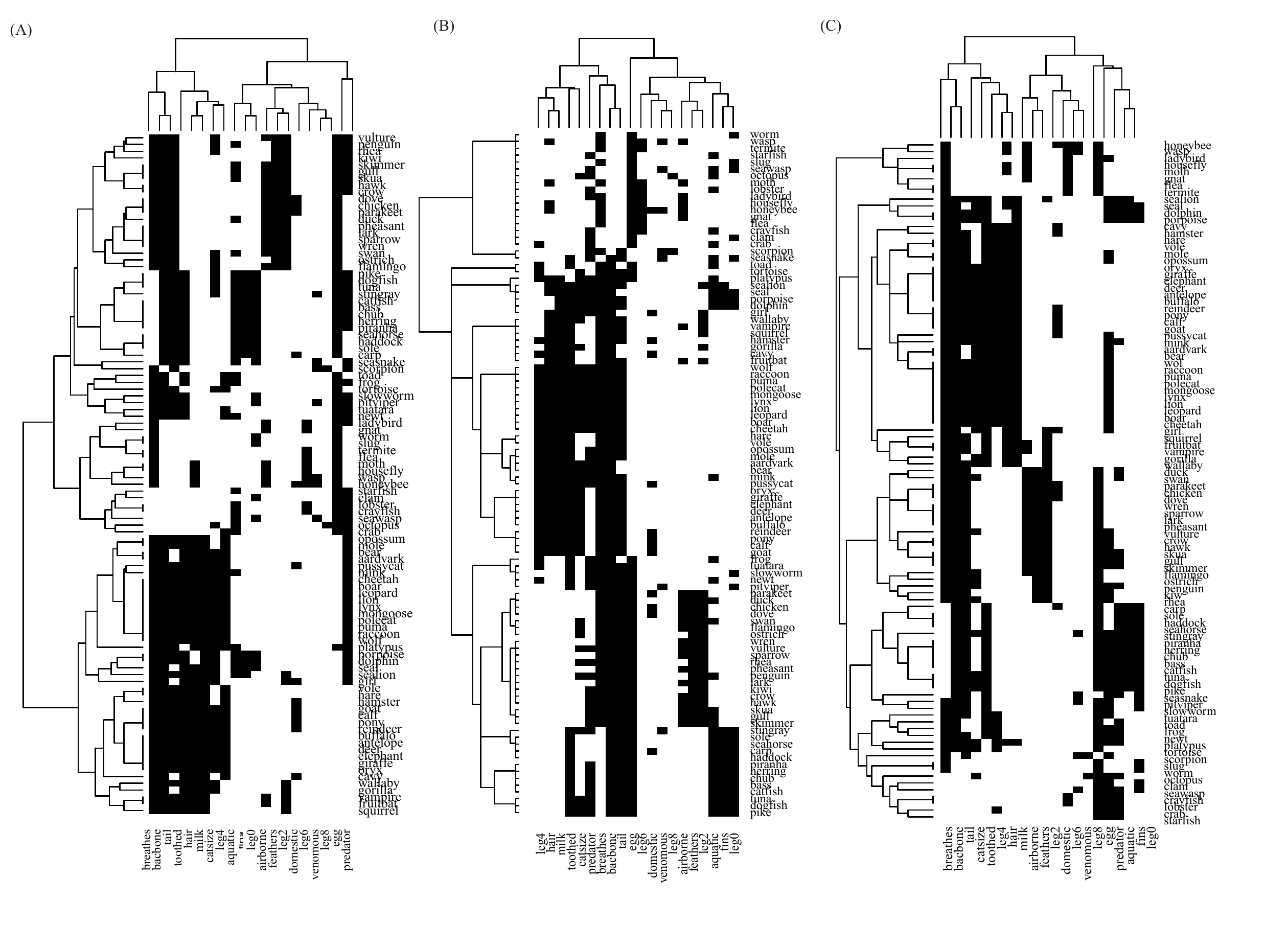}
	\caption{ Resultant heatmaps.(A)DM with UPGMA vs UPGMA (B) DM with DCG-vs-HC; (C) HC-vs-HC}
	\label{fig:dcg-vs-hcheatmap}
\end{figure}

From Data Mechanics computations, three resultant heatmaps superimposed with a UPGMA (HC with average module), DCG tree and HC-tree (with complete module) on the species-axis and HC tree on the feature-axis are respectively given in three panels of Fig.~\ref{fig:dcg-vs-hcheatmap}. To achieve ``uniformity" in blocks, we update distance with respect to a tree structure along the counterpart axis in each iteration. Specifically extra dimensions of averaging with respect to each cluster on the counterpart axis are taken into account. Hence it is necessary to emphasized that the HC tree on species axis of panel (A) is not the commonly used one via UPGMA method, which is a HC tree without going through the iterating procedure of Data Mechanics.

The three HC trees on the feature axis of 20 features of the three panels basically are only slightly different. Two detailed differences are seen. The egg feature is kind of an outlier among the 20 features in panels (A) and, particularly in panel (B), while it is not in panel (C). The predator feature in panel (B) is grouped with features of backbone, breathes, toothed, leg-4, while this feature is grouped with egg feature to form an isolated branch in panel (A), and with features of egg, aquatic, fine, leg-0, in panel (C). Its location in panel (B) seems to make more biological senses than that in panels (A) and (C).

\begin{center}
	\begin{figure}[hbtp]
		\centering
		\includegraphics[scale=0.6]{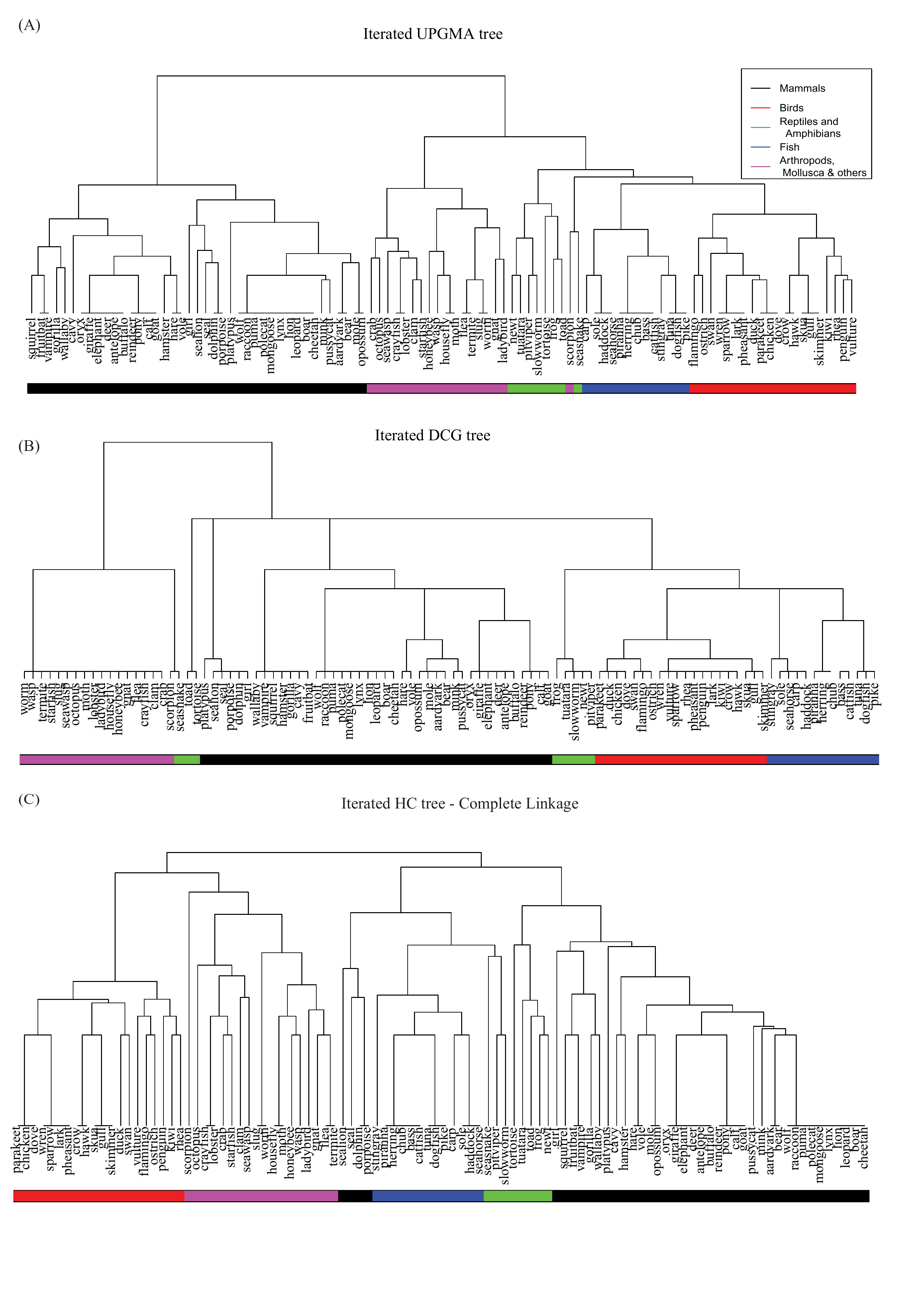}
		\caption{Comparison of the three final trees of Zoo data.}
		\label{fig:zootreecomparison3}
	\end{figure}
\end{center}

As for the three trees on species-axis of the three heatmaps are not drastically distinct, see more details through Fig.~\ref{fig:zootreecomparison3}. The DCG tree in panel (B) of Fig. ~\ref{fig:zootreecomparison3} is equipped with 8 tree-levels, from the top to bottom, composing of $\{2, 5,  8, 10, 12, 14, 16, 22\}$ branches, while the HC based trees in panels (A) and (C) of Fig.~\ref{fig:zootreecomparison3} have many tree levels, which compose of successive number of branches: from 2 to many. At the level of 2 branches of DCG tree, one big branch and one small one. The members of the small branch is consisting two core clusters: one have 17 members: worm, slug, seawasp, wasp,...,  and the other has a tiny core cluster of two: {scorpion, seasnake}. The members of this small branch of DCG tree is further bifurcated many times in the UPGMA and HC trees. Such multiple bifurcations on a core cluster of DCG tree is not supported by data based on the energy densities in Fig.~\ref{fig:energyzoocomparison3}. This is one of their major differences among the three trees. Minor difference is seen specifically through the branch of {scorpion, seasnake}. This tiny branch is located next to two big branches of Bird and Fish in UPGMA, while they are completely separated belonging to two major branches at the 2-branch level of the HC tree. These separations seem unnatural biologically.

Beyond aforementioned structural differences among the UPGMA, HC and DCG trees, other major differences are visible on branching structures above the bottom-level of DCG tree. For instance on the 5-cluster level of DCG tree, the {girl, dolphin, porpoise, seal, sealion} is a core cluster. It and {platypus} form a branch on the 16-cluster level of DCG, and separates from the rest of mammals. This small branch is broken down in UPGMA, even more so in HC tree. Another core cluster of mammals: {wallaby, vampire, squirrel, hamster, gorilla, cavy, fruitbat}, is also taken apart in both UPGMA and HC trees. We think these cases are evidence that tree structures based on HC algorithm are not coherently supported by data. This conclusion is somehow visible from the heatmaps in Fig.~\ref{fig:dcg-vs-hcheatmap} via the intuitive concept of uniformity on matrix lattice.

\subsection{Plasmodium Data}

 Next we use a neucleotide sequence data on the Malaria parasite Plasmodium. The $11 \times 221$ matrix of aligned nucleotide sequences of  the 11 plasmodium isolates were taken from \cite{Efron}. This data originally was a $11 \times 1620$ matrix of aligned nucleotides A, G, T, C. For analysis only $221$ polytypic sites were extracted from this matrix. For computation purpose A,G,C, and T was digitally coded by 1,2,5 and 6 respectively. The 11 species of Plasmodium came from different animals as described below.
\begin{center}
\begin{tabular}{|c|c|c|c|c|c|c|c|c|c|c|}
	\hline
	1 & 2 & 3 & 4 & 5 & 6 & 7 & 8 & 9 & 10 & 11 \\
	\hline
	Pre & Pme  & Pma & Pfa & Pbe & Plo & Pfr & Pkn & Pcy & Pvi & Pga \\
	\hline
	Chimp & Lizard & Human & Human & Rodent & Bird & Monkey & Monkey & Monkey & Human & Bird \\
	\hline
\end{tabular}
\end{center}

\begin{figure}[hbtp]
	\centering
	\includegraphics[scale=0.5]{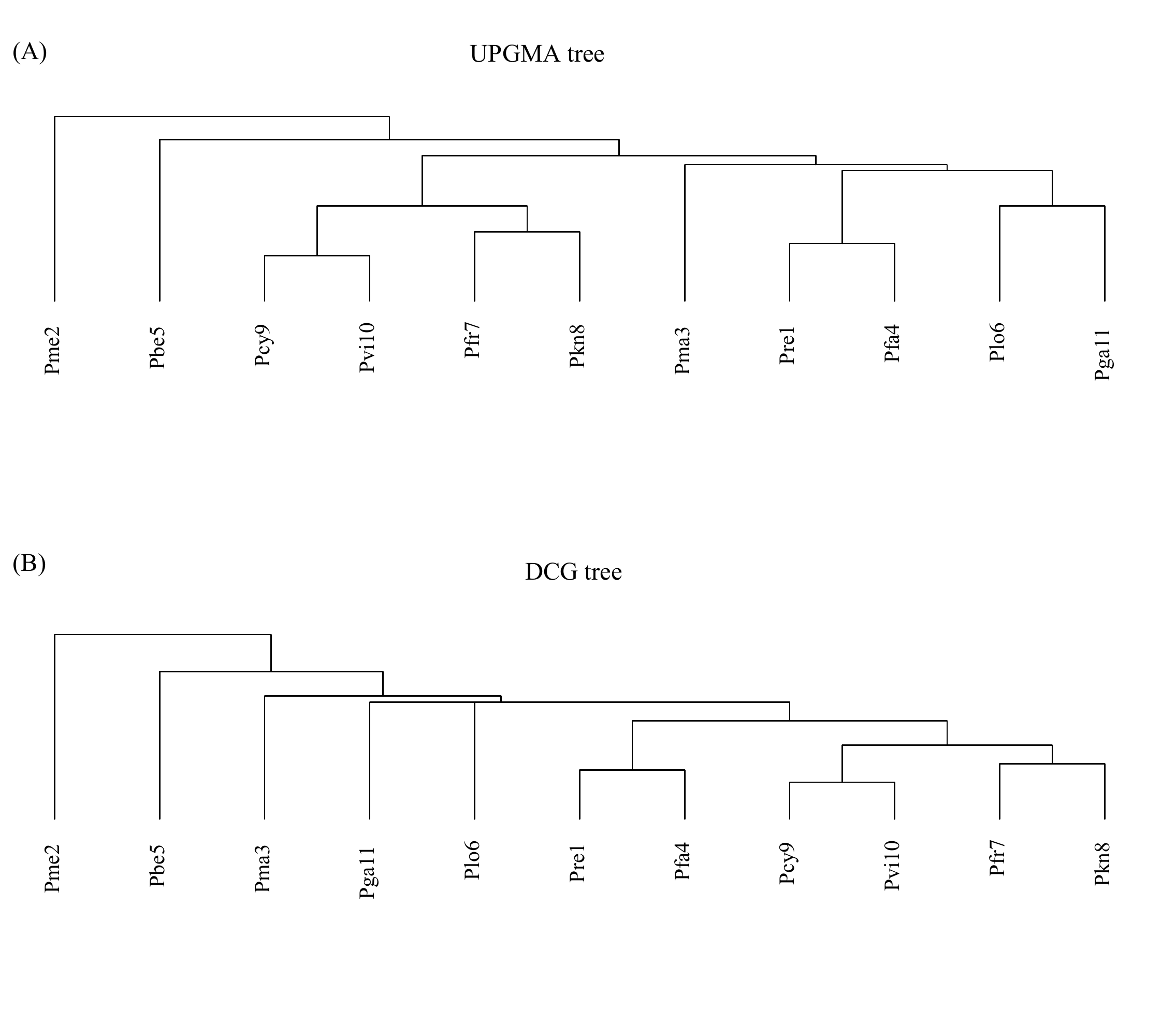}
	\caption{Comparison of UPGMA and DCG trees on the 11 species of plasmodium.}
	\label{fig:plasmodiumtwotrees}
\end{figure}

Based on the numeric representation of the matrix, a DCG tree was built and compared with an UPGMA tree in figure \ref{fig:plasmodiumtwotrees}.
In the DCG tree, the core cluster {Pre1, Pfa4} coupling with the core cluster {Pcy9, Pvi10} and cluster {pfr7, pkn8} becomes a cluster, while it is separated the two clusters in the UPGMA tree. This is the major difference between the two trees.

\begin{figure}[h]
	\centering
	\includegraphics[width=0.8\linewidth]{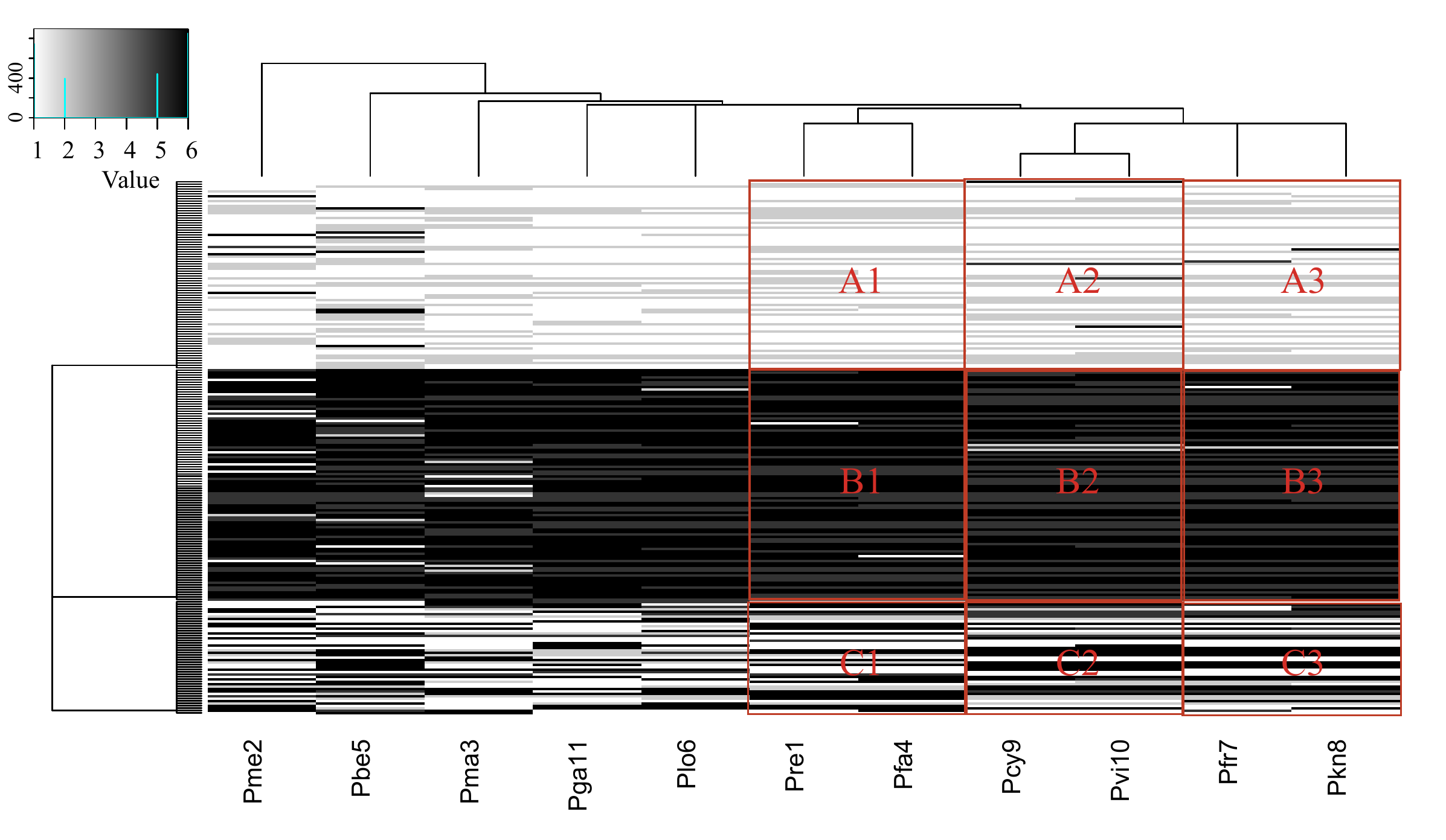}
	\caption{Heatmap showing 9 basic blocks discovered through the Datamechanics on the Plasmodium data. Columns are showing 11 species of Plasmodium and rows are showing the A,G,T,C coding. 1 = A, 2 = G, C = 5, T= 6.}
	\label{fig:plasmodiumdcgheatmapblocks}

\end{figure}

Next we conduct the matrix mimicking technique to understand the structure of the DCG tree on the species. The figure \ref{fig:plasmodiumdcgheatmapblocks} shows a permutation of the transpose of the original matrix with the two DCG trees imposed, one on the species(columns) and the other on the sites(rows). The numbers 1, 2, 5 and 6 are represented by white, two shades of gray and black. The DCG tree on the sites show three different types of homogeneous blocks, and the DCG tree on the species shows us three core clusters containing (Pre1, Pfa4), (Pcy9, Pvi10)  and (Pfr7, Pkn8). The rest of the species are probably quite different from these core clusters, since in the DCG tree they are arranged in outlying branches. Thus using the two trees, we can divide the permuted matrix in 9 core blocks, which are clearly homogeneous in nature. The 9 core blocks are marked and named in figure \ref{fig:plasmodiumdcgheatmapblocks}.

\begin{figure}
	\centering
	\includegraphics[scale=0.7]{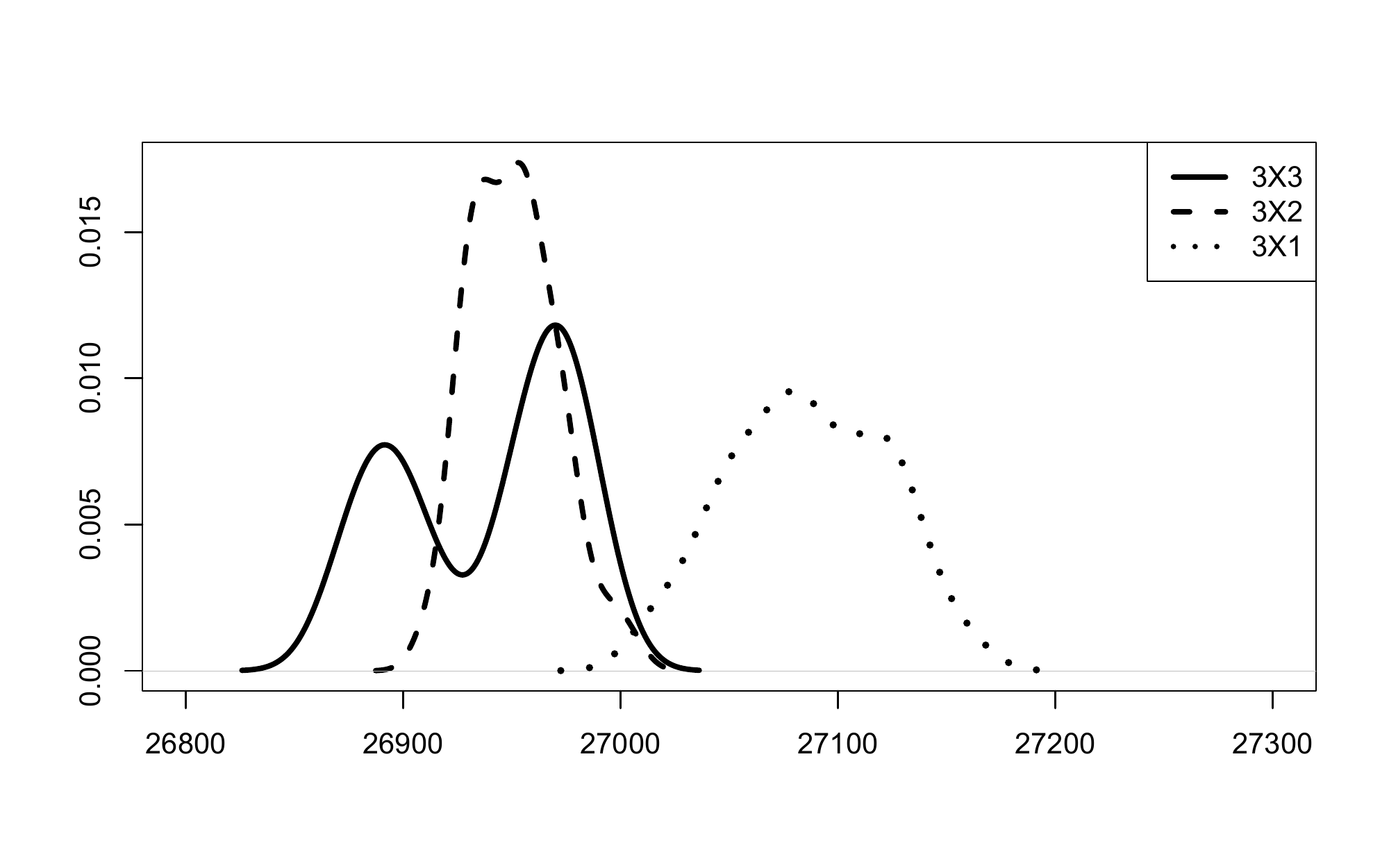}
	\caption{Plasmodium Energy Comparison}
	\label{fig:plasmodiumenergy}
\end{figure}

To understand the tree structure of the DCG tree next we mimic each of the 6 blocks individually (100 times each) and combine them to get 100 mimicked version of the permuted matrix. For each the mimicked version, we find the energy of the matrix and plot them as a solid density line in figure \ref{fig:plasmodiumenergy}. Next we go up the DCG tree one level and merge two blocks at a time and get  core blocks: A1, (A2:A3), B1, (B2:B3), C1 and (C2:C3). In this $3\times 2$ setup we again repeat the mimicking process and plot the energies as a dashed density line in figure \ref{fig:plasmodiumenergy}. In the next step we combine all A blocks, all B blocks, and all C blocks to get a $3\times 1$ setup for mimicking. This time the energies are plotted in the same figure as a dotted line. The mimicking algorithm for an (A, G, C, T) matrix can be found in Appendix A.

Clearly as we merge more blocks that are homogeneous inside but pretty much heterogeneous among themselves, we keep getting higher density patterns. This means the two merging of nodes in these two levels of the DCG tree are viable as the blocks are substantially different. The $3 \times 3$ setting might not be very different from the $3\times 2$ setup, however the $3 \times 1$ setup is a lot different than these two. That indicates that the core cluster of Pcy9 and Pvi10 is probably somewhat similar to Pfr7 and Pkn8, although the core cluster of Pre1 and Pfa4 are much different than these four species clusters.

\subsection{Cryptosporidium Data}
[Data description] The second genomic data on which we apply our techniques are DNA sequences of the parasite Cryptosporidium, that causes diarrhea in mammals. There are many existing species of this parasite, and the humans are mostly affected by C.parvum and C. hominis and sometimes by C. canis, C. felis, C. meleagridis, and C. muris. In our study DNA sequences from a lot of different strains of the parasite were gathered, aligned, and cleaned to form a gapped $\{ A, G, C, T\}$ matrix.

\begin{figure}[hbtp]
	\centering
	\includegraphics[scale=0.8]{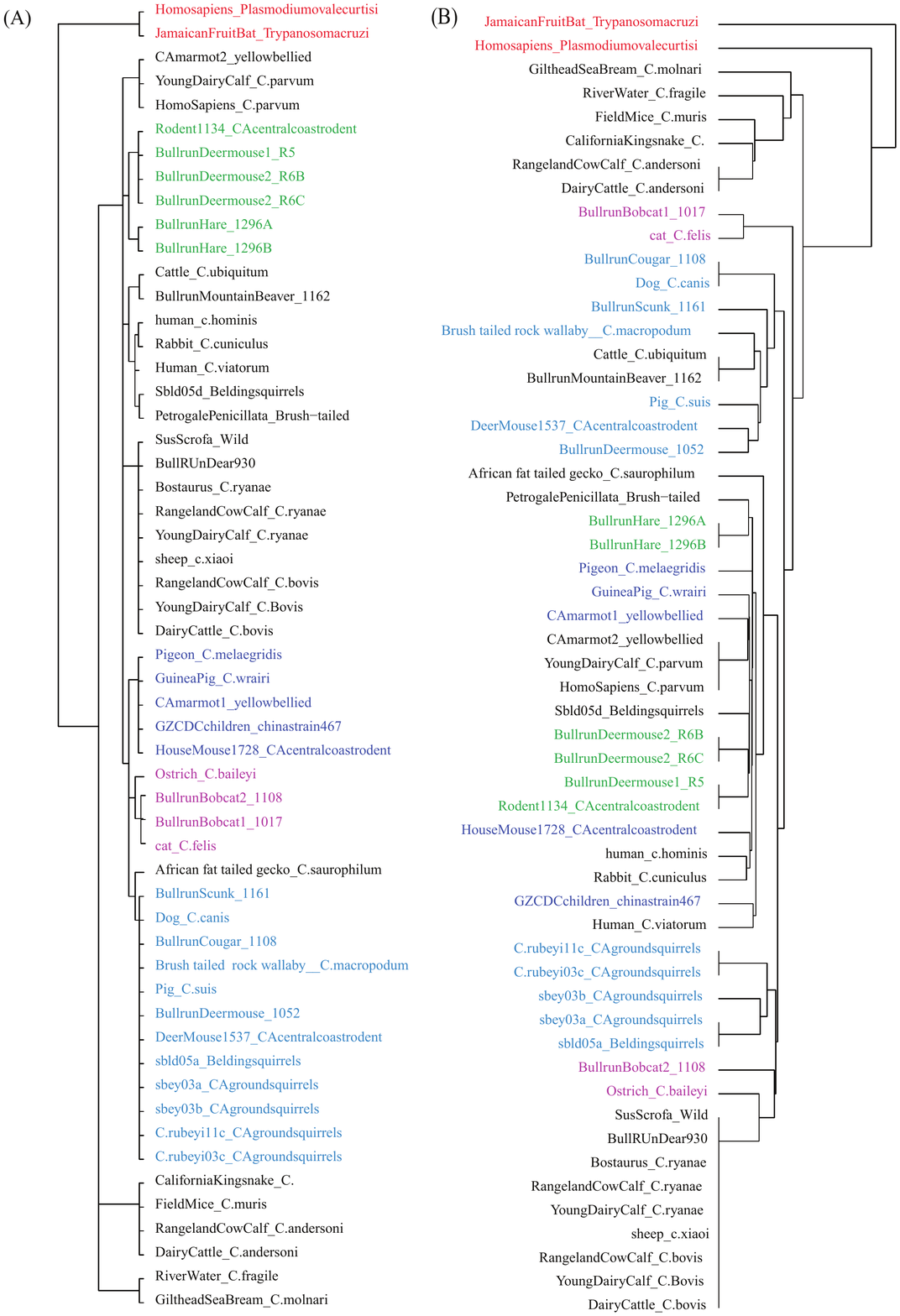}
	\caption{Panel A: DCG tree , Panel B: UPGMA tree. }
	\label{fig:dcg-vs-upgma-criptovertical}
\end{figure}

In the data two different bacteria namely, Plasmodium and Trypanosoma cruzi (marked by red in figure \ref{fig:dcg-vs-upgma-criptovertical}) were deliberately mixed in the data to understand their similarity or dissimilarity with Cryptosporidium. Unlike the Plasmodium data, we kept the gaps before constructing the DCG tree, and used a scoring mechanism (described in appendix B) to find the similarity of the aligned sequences instead of simply representing the four bases by numbers. This method allows us to retain a lot more information and also help us avoid the bias due to the choice of the representative numbers of the bases A, G, C and T.

Next based on the similarity matrix constructed as such, a DCG and an UPGMA tree was built. They are compared in figure \ref{fig:dcg-vs-upgma-criptovertical}. To compare the two trees some important core clusters of the DCG tree had been colored, and the redistribution of those colored leaves are shown in the UPGMA tree on the right. The magenta leaves showing cat, two bobcats and ostrich are together in the DCG tree, however in the UPGMA tree the two bobcats get separated in far apart branches. The light-blue branch consisting most ground squirrels, two mouse, pig, cougar, dog, scunk etc are so similar that DCG tree do not divide them in further branches. However the UPGMA tree separates them in three different branches with quite a big distance between the branches.

To understand whether this branching is viable we extract only the animals colored by light-blue, re-score their sequences, and then make a DCG tree on only those species. In the appendix in the right panel of the  figure \ref{dcg-heatmap-crypto}, we see an almost similar branching on the animals. This indicates that the sequences from these animals are different from each other, but during multiple comparison with all other sequences, the dissimilarity is probably not striking, and hence the DCG do not separate them into more branches.

The green cluster in DCG tree is separated in two branches in the UPGMA, and the deep-blue branch made of pigeon, guinea pig, human child, marmot and house mouse splits in 3 different parts in UPGMA tree. Note that both the trees fail to split the branch of Dairy cows (black) and sheep. To make new branches out of this group we extracted them from the whole group, re-aligned and re-scored the 9 sequences and ,add a DCG tree on them again. The new tree on this subsection can be seen in figure \ref{hc-cow-squirrel} in the Appendix B.

\section{Conclusion}

It is known fact that given a set of organisms, any Phylogenetic tree built on them is a hypothesis in nature. It is very difficult to know what is the true hierarchical relationship, or the evolutionary pathway in those organisms. Hence instead of comparing the different existing methods of building Phylogenetic trees, our effort was to establish Data Cloud Geometry as a rigorous non-parametric, distance based alternative approach to Phylogenetics. The DCG trees on Zoo, Plasmodium, and Cryptosporidium data were built and shown to be very reasonable comparing to the widely used UPGMA trees. Further, we have tried to address the over-splitting issue of the traditional bifurcating Phylogenetic trees, and established a method of checking the viability of tree levels. This method could be applied on not only the multi-furcating DCG trees, but also on any traditional Phylogenetic methods. The merit of the DCG based Phylogenetic tree lies in the fact that it has the potential to increase and decrease the number of levels of the tree, depending on need and context, by choosing the proper tuning parameters called temperatures, and the number of such temperatures selected dictate the detail of the levels of the tree.

Being a distance based non-parametric method, DCG trees are free from the unnecessary assumptions of the traditional Cladistic methods of Phylogenetics, and at the same time much more rigorous than the common Phenetic methods. The computational complexity is not too high and depends on the number of scales through which we re-explore the basic distance matrix. At each scale the DCG mechanism re examines the relationship of the nodes through the node-removal technique of the regulated Markov chain algorithm, and that is especially important since, before finalizing the tree we can check the viability of the selected scales by the matrix mimicking technique.

We must acknowledge that the matrix mimicking technique is still in its infancy and right now only applicable to binary and gene sequence type categorical matrices. For continuous data, we could still apply appropriate categorization techniques \cite{hsiehroy}, and carry out the nested binary slicing technique multiple times.

\newpage
\section*{Appendix}
\appendix

\section{ Algorithm for mimicking $\{A, G, C, T\}$ Matrix}
\renewcommand\thefigure{\thesection.\arabic{figure}}
The following algorithm is for mimicking an $\{A, G, C, T\}$ marix with a given marginal distribution of the 4 bases A, G, C and T. The idea is to apply the Binary slicing procedure in a nested manner. We first slice the bases data in two parts, one containing A or G and the other containing C or T. After randomly generating the placeholders of these two categories from their marginal distribution, we go inside each part and find the placeholders of A out of all places for A and G, and similarly for C out of all places for C and T. The version A describes this algorithm.

However if the matrix to be mimicked is large or very heterogeneous, then often the version A of the algorithm takes a lot of time to find a solution. Often the first half of this nested algorithm finds such places for $\{A, G\}$ and $\{T, C\}$,  that finding a solution in the second step becomes almost impossible. Thus for large matrices we modify this algorithm and do not allow the positions of $\{A, G\}$ and $\{T, C\}$ to be exchangeable. This forces us to allocate more memory for storage of the input of the algorithms, but improves the computation speed a lot. The modified algorithm is explained in version B.

\begin{enumerate}
	\item[A.]      [Algorithm for non-exchangeable position between $ \{A, G\}$ and $\{T, C\}$ ]
	\begin{enumerate}
		
		\item[1. ]      Denote the block by B. And let $B_{A, G}$ be the block showing only entries {A} or {G}, while all entries of occupied by {T} or {C} being marked and not being allowed to be changed. Similarly  define $B_{T, C}$;
		
		\item[2.]       Encoding {A} to be 0, and {G} to be 1, so $B_{A, G}$ is converted in $B_{0, 1|A, G}$. Likewise converting   $B_{T, C} $into $B_{0,1|T, C}$;
		
		\item[3.]       Apply Binary Slicing algorithm on $B_{0, 1|A, G} $ and $B_{0,1|T, C}$, separately.
		
		\item[4.]       Put together simulated $B^*_{0, 1|A, G} $ and $B^*_{0,1|T, C}$ into a simulated block $B^*$.
		
	\end{enumerate}

	\item[B.]      [Algorithm for exchangeable position between $ \{A, G\}$ and $ \{T, C\}$]
	\begin{enumerate}
		\item[1.]       Encoding {A, G} to be 0, and {T, C} to be 1, so B is converted in binary block B[0,1].
		
		\item[2.]       Apply Binary Slicing algorithm to simulate a new binary block $B^*[0,1]$;
		
		\item[3.]       Let $B^*[0,1]_{0}$ be the block showing only entries {0}, while all entries of occupied by {1}  being marked and not being allowed to be changed. Similar defined $B^*[0,1]_{1}$.
		
		\item[4.]       Apply Binary Slicing algorithm on $B^*[0,1]_{0}$ by subject to row and column {G}-degree sequences of {G} calculated from B. Denote the simulated block as $B^{**}[0,1]_{A, G}$;
		
		\item[5.]       Apply Binary Slicing algorithm on $B^*[0,1]_{1}$ by subject to row and column {C}-degree sequences of {C} calculated from B. Denote the simulated block as $B^{**}[0,1]_{T, C}$;
		
		\item[6.]       Put together $B^{**}[0,1]_{A, G}$  and $B^{**}[0,1]_{T, C}$ into a simulated block $B^{**}$.
	\end{enumerate}
	
\end{enumerate}

For the Plasmodium data we have used version A for generating all blocks in the $3 \times 3$ and $3 \times 2$ setup. However the combined matrices were large in the $3 \times 1$ setup, so we used the version B for higher computation speed. The original vs mimicked matrices in the $3 \times 2$ split is illustrated in figure \ref{fig:bootblocksplasmodium}.

\setcounter{figure}{0}
\begin{figure}[hbtp]
	\centering
	\includegraphics[scale=0.5]{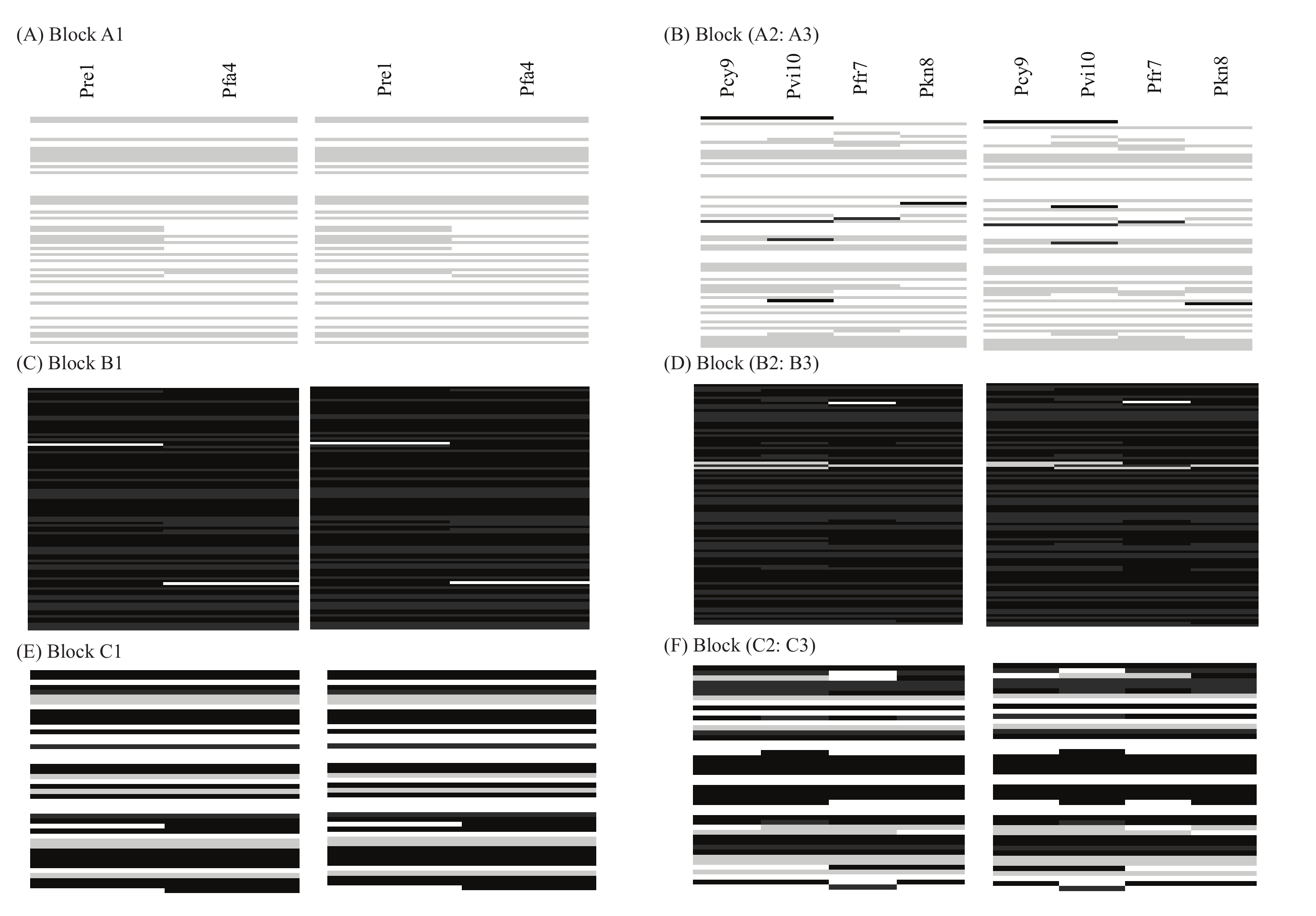}
	\caption{Original and mimicked blocks in the $3\times 2$ setup. In each panel left matrix is original and the right matrix is mimicked according to algorithm version A.}
	\label{fig:bootblocksplasmodium}
\end{figure}

\section{Alignment and Scoring method for Cryptosporidium}
\subsection{Step 1: Alignment}
We are given with 53 sequences of Cryptosporidium collected from different hosts and 2 other sequences.The first step is using a software to align the sequences. Now we have 55 strings of aligned sequences, with a lot of gaps. The next and most important step is scoring the sequence strings so that we get a score for each pair of them indicating the similarity of the strings. The following picture shows a section of the aligned sequences. The multiple sequence alignment was done using CLUSTAL X \cite{clustalx} software.

\setcounter{figure}{0}
\begin{figure}[htbp]
	\begin{center}
		\includegraphics[scale=0.65]{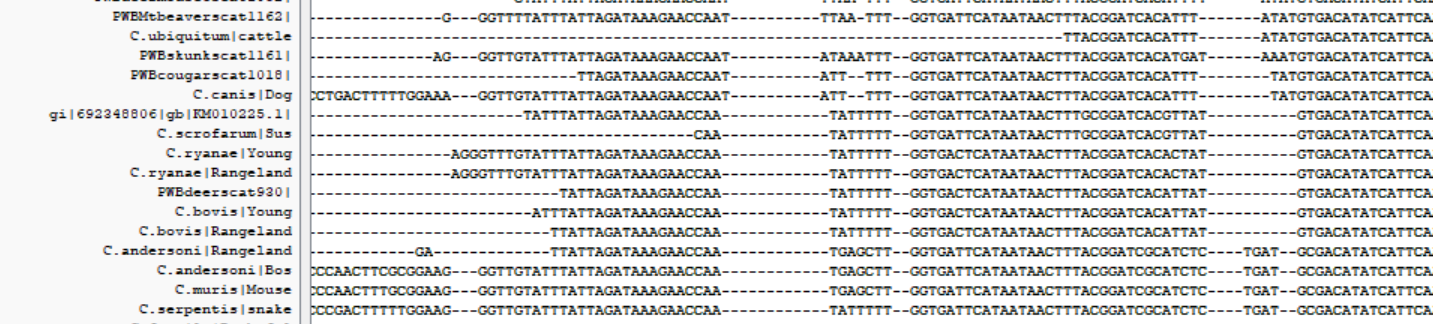}
		\caption{Example of aligned sequences from CLUSTAL X.}
		\label{aln-crypto}
	\end{center}
\end{figure}

\subsection{Step 2: Scoring method}

\begin{enumerate}
	\item Let A be the gap opening penalty and let B be the gap extension penalty. In this format the total penalty for a gap of length L should be $A+B\times L.$  When two lines are paired if one of them has the gaps and other one does not, then we subtract a penalty of $A+B\times L$ from the score of the pair. If both of the sequences has  continued gaps at the same places then we do not penalize them at all.
	
	\item Also we do not count the gaps for penalizing when one sequence has already started and the other did not, or one sequence has ended but the other did not. For example in the above picture while scoring the pair sbey03a and sbey03b a we see that sbey03a ends before the other and while the other keeps continuing. sbey03a continues with gaps. For this pair we will consider only till the position where the shorter sequence sbey03a ends, and discard the rest from the scoring process.

	\item We add 1 to the score for each matching pair and do not add or subtract anything for a mismatching pair which does not include a gap.
	
	\item After we sum up all the positive scores from the matching base pairs and subtract all the gap penalties, we divide the score by  a common length(the total number of positions/places under comparison in that particular pair of sequences. Essentially we do not count the both gap places in this length). After this standardization, we subtract all the scores by the minimum and divide by the maximum so that after the process we get a score between 0 and 1, inclusive.
	
\end{enumerate}

\subsubsection*{An example of scoring}

If we have two sequences like below \\
A- - - G- - - - TTCA- - - - - \\
A-T TC- - - - TTCGATG- - \\

Then the scoring will be like this.For matching A, T, T, C the score will get 4.  There is just one place for penalty, the two gaps and two T's which gets the the penalty $A+2B$. hence the score becomes 4-(A+2*B).

\begin{center}
\begin{figure}[htbp]
		\includegraphics[scale=0.6]{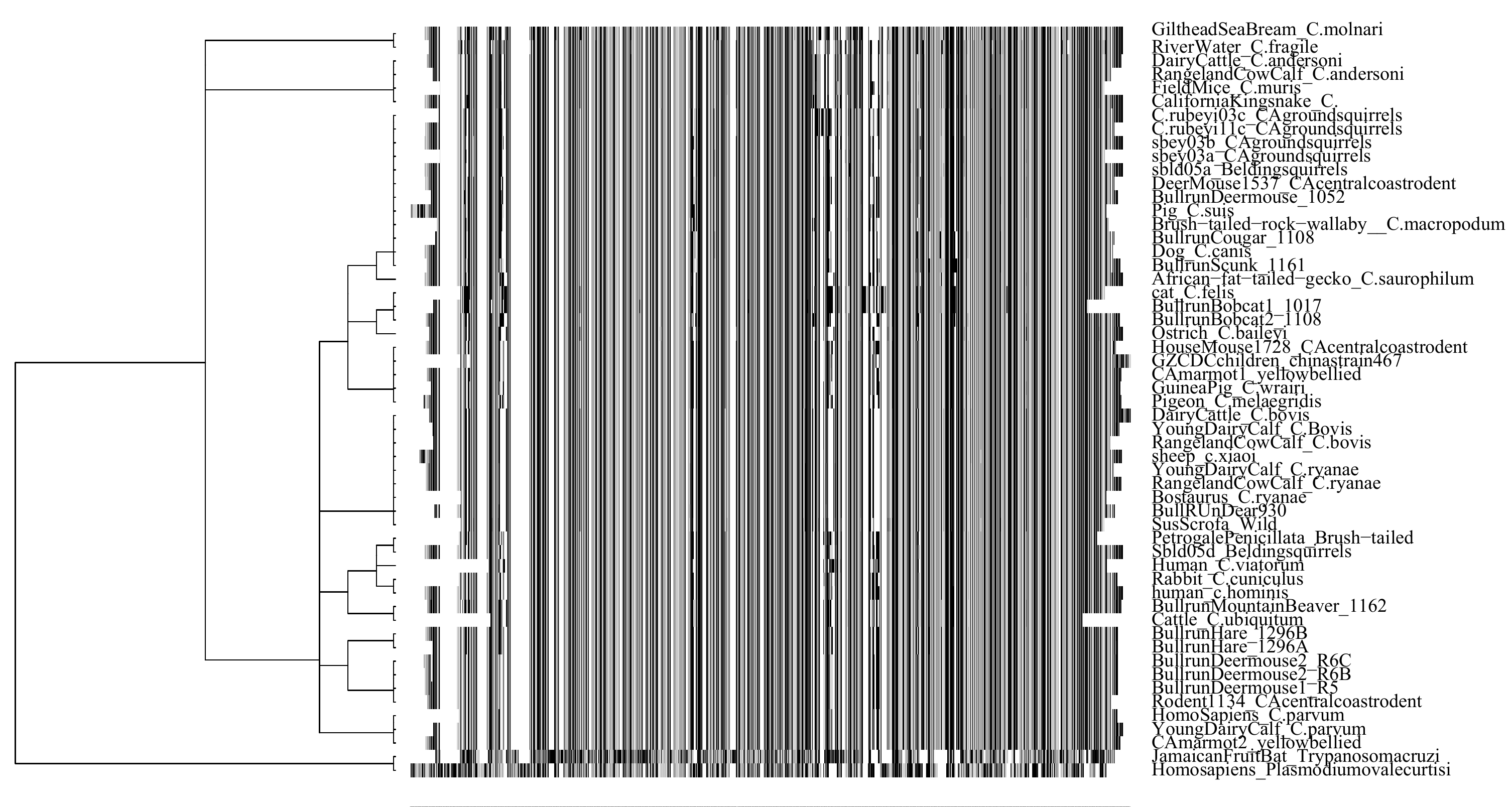}
		\caption{DCG tree on all the species permuting the aligned DNA sequence matrix. Here white  represents gaps, and shades of gray upto black represents 1, 2, 5, and 6 corresponding to A, G, C and T respectively. }
		\label{dcg-heatmap-crypto}
\end{figure}
\end{center}

\subsubsection{Step 3: Similarity matrix construction}
At the final step we create a similarity matrix $S$ where $S_{i,j}$ is the standardized score found after the whole scoring procedure is done. Using this similarity matrix we can proceed towards the DCG tree construction.

\subsubsection{DCG tree constructed}

\begin{center}
\begin{figure}[htbp]
\includegraphics[scale=0.4]{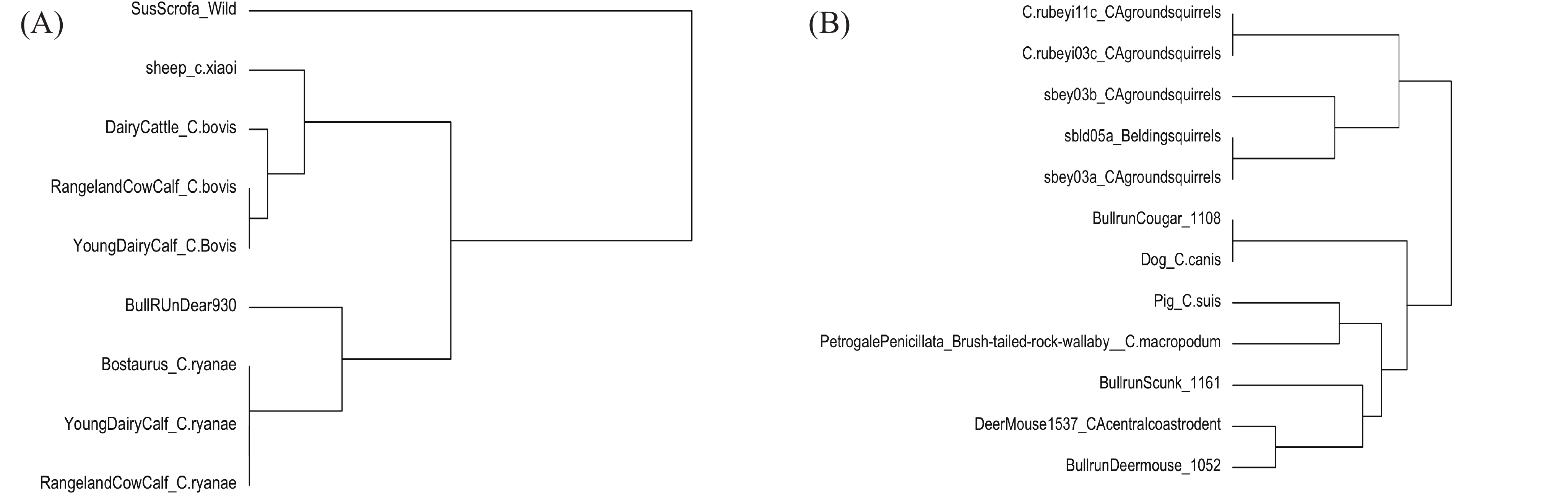}
\caption{New DCG trees on Dairy calf section and Squirrels section from figure \ref{fig:dcg-vs-upgma-criptovertical} after they are extracted and re-scored.}
\label{hc-cow-squirrel}
\end{figure}
\end{center}

To do the scoring we took A=15 and B=0.2. The DCG tree constructed is given in figure \ref{dcg-heatmap-crypto}. However because of ignoring most gaps in the scoring procedure the tree has two large branches in which further branching is not possible. This will be easy to correct if we take those homogeneous two groups separately and reconstruct the scoring matrices and the trees. The finer branching in those two branches are given in figure \ref{hc-cow-squirrel}.

\end{document}